\documentclass[preprint,12pt,authoryear]{elsarticle}

\usepackage{amssymb}
\usepackage{amsmath}
\usepackage{tabularx}
\usepackage{booktabs}
\usepackage{changepage}
\usepackage{url}
\usepackage{array}
\usepackage{float}
\usepackage[colorlinks=true,linkcolor=blue,citecolor=blue,urlcolor=blue]{hyperref}
\journal{Epidemics}

\newcolumntype{Y}{>{\centering\arraybackslash}p{1.8cm}}

\begin{document}

\begin{frontmatter}

%% Title, authors and addresses

%% use the tnoteref command within \title for footnotes;
%% use the tnotetext command for theassociated footnote;
%% use the fnref command within \author or \affiliation for footnotes;
%% use the fntext command for theassociated footnote;
%% use the corref command within \author for corresponding author footnotes;
%% use the cortext command for theassociated footnote;
%% use the ead command for the email address,
%% and the form \ead[url] for the home page:
%% \title{Title\tnoteref{label1}}
%% \tnotetext[label1]{}
%% \author{Name\corref{cor1}\fnref{label2}}
%% \ead{email address}
%% \ead[url]{home page}
%% \fntext[label2]{}
%% \cortext[cor1]{}
%% \affiliation{organization={},
%%            addressline={}, 
%%            city={},
%%            postcode={}, 
%%            state={},
%%            country={}}
%% \fntext[label3]{}

\title{Dual-Model Framework for CHIKV Transmission Modeling: ODE and Petri Net Analysis of the 2025 Foshan Outbreak} %% Article title

% use optional labels to link authors explicitly to addresses:
% \author[label1,label2]{}
% \affiliation[label1]{organization={},
%             addressline={},
%             city={},
%             postcode={},
%             state={},
%             country={}}
%
% \affiliation[label2]{organization={},
%             addressline={},
%             city={},
%             postcode={},
%             state={},
%             country={}}

%% link authors to affiliations via optional labels
\author[aff1,aff2,aff3]{Hong Liu\fnref{equal}}
\author[aff1,aff2,aff3]{Jingjing Tian\fnref{equal}}
\author[aff6]{Yiping Li}
\author[aff7]{Yuyang Zhong}
\author[aff1,aff2,aff3]{Haibo Yang}
\author[aff1]{Dong Chen}
\author[aff1,aff2,aff3,aff4,aff5]{Zifeng Yang\corref{cor1}}

\fntext[equal]{These authors contributed equally to this work.}
\cortext[cor1]{Corresponding author.}
\ead{jeffyah@163.com}

%% Author affiliations
\affiliation[aff1]{organization={Respiratory Disease AI Laboratory in Epidemic Intelligence and Applications of Medical Big Data Instruments, Macau University of Science and Technology},
            addressline={}, 
            city={Taipa},
            postcode={}, 
            state={Macau SAR},
            country={China}}
\affiliation[aff2]{organization={Faculty of Innovation Engineering, Macau University of Science and Technology},
            addressline={}, 
            city={Taipa},
            postcode={}, 
            state={Macau SAR},
            country={China}}
\affiliation[aff3]{organization={Institute of Systems Engineering, Macau University of Science and Technology},
            addressline={}, 
            city={Taipa},
            postcode={}, 
            state={Macau SAR},
            country={China}}
\affiliation[aff4]{organization={State Key Laboratory of Respiratory Disease, National Clinical Research Center for Respiratory Disease, Guangzhou Institute of Respiratory Health},
            addressline={}, 
            city={Guangzhou},
            postcode={}, 
            state={Guangdong},
            country={China}}
\affiliation[aff5]{organization={Guangzhou National Laboratory},
            addressline={}, 
            city={Guangzhou},
            postcode={}, 
            state={Guangdong},
            country={China}}
\affiliation[aff6]{organization={Faculty of Humanities and Arts, Macau University of Science and Technology},
            addressline={}, 
            city={Macau},
            postcode={999078}, 
            state={Macau SAR},
            country={China}}
\affiliation[aff7]{organization={China (Hong Kong) English School},
            addressline={}, 
            city={Zhongshan},
            postcode={}, 
            state={Guangdong},
            country={China}}

%% Abstract
\begin{abstract}
This study constructs a dual-model framework integrating Ordinary Differential Equations (ODE) and Petri Nets (PN) to analyze the 2025 Chikungunya outbreak in Foshan City, China. We employ SEICR compartmental modeling to compare two distinct approaches under identical epidemiological scenarios and evaluate intervention effectiveness through three-phase fitting protocols. Both models demonstrate excellent accuracy with MAE of 18.77-18.91 cases and RMSE of 36.52-36.54 cases. Models predicted epidemic peaks at day 32 (406 cases), 3 days earlier than observed (day 35, 432 cases), with 6.0\% peak value error. Reproduction number analysis revealed initial $R_0$ of 14.67 (ODE)/13.90 (PN), with effective reproduction numbers decreasing through intervention phases: 7.85/7.86 after Phase 1, 7.59/7.56 after Phase 2, and 0.059 in Phase 3, achieving transmission blockade. Sensitivity analysis showed recovery rate ($\gamma$) as the most sensitive parameter (Sobol index 0.9672), explaining 96.72\% of $R_0$ variation. This study presents the first systematic ODE-Petri Net comparison, providing a novel dual-model framework for vector-borne disease modeling with significant theoretical and practical value for epidemic control strategy formulation.
\end{abstract}

% %%Graphical abstract
% \begin{graphicalabstract}
% %\includegraphics{grabs}
% \end{graphicalabstract}

%%Research highlights
\begin{highlights}
\item Systematic numerical comparison between ODE and Petri Net models for CHIKV transmission
\item Recovery rate identified as most sensitive parameter (Sobol index: 0.9672) for epidemic control
\item Three-phase intervention analysis shows effective transmission blockade (\(R_{\mathrm{eff}}\): 0.059)
\item Novel methodological framework for vector-borne disease modeling and control strategy
\end{highlights}

%% Keywords

\end{frontmatter}

%% Add \usepackage{lineno} before \begin{document} and uncomment 
%% following line to enable line numbers
%\linenumbers

%% main text
%%

%% Use \section commands to start a section
\section{Introduction}
\label{sec:introduction}

Chikungunya virus (CHIKV) is a public health concern in tropical and subtropical regions, with recent outbreaks demonstrating the need for rapid response and accurate transmission modeling. In July 2025, a significant CHIKV outbreak occurred in Foshan City, Guangdong Province, China, originating from an imported case detected through active surveillance in Shunde District on July 8th \citep{zhang2025foshan}. The outbreak rapidly expanded, with 1,873 confirmed cases reported by July 19th, primarily concentrated in Shunde District (1,790 cases) across Letong, Beijiao, and Chencun towns, with additional cases in Chancheng (49 cases) and Nanhai (34 cases) districts \citep{wan2025viral}. All cases were mild with no severe or fatal outcomes, and 720 patients had recovered by that time. The outbreak continued to grow, reaching 2,934 cases in Shunde District by July 22nd, and even spread beyond Guangdong Province to Macau Special Administrative Region \citep{zhang2025foshan}. Timely estimation of the basic reproduction number (\(R_0\)) and the phase-specific effective reproduction number (\(R_{\mathrm{eff}}\)) is central to situational awareness and control evaluation during such outbreaks. Recent epidemiological studies of CHIKV outbreaks have demonstrated the importance of timely R estimation for intervention planning \citep{zhang2025foshan,wan2025viral}.Traditional CHIKV modeling has relied heavily on compartmental ODE models for transmission dynamics and reproduction number estimation \citep{vanden2002reproduction,martcheva2015introduction}. However, existing ODE-based approaches face limitations in representing complex intervention scenarios, resource constraints, and the discrete nature of disease transmission events \citep{connolly2022from}.Compartmental ordinary differential equation (ODE) models are standard; however, Petri Nets (PN) provide a discrete-event, compositional alternative well-suited for switching interventions and resource-aware processes.

Petri Nets have seen growing use as a complementary formalism to ODEs for epidemiological modeling and system composition \citep{segovia2025petri,connolly2022from}. Generalized and colored/stochastic variants enable concurrency and resource-aware representations \citep{peng2021gspn}. In vector-borne contexts, MGDrivE employs stochastic Petri Nets for mosquito life history and epidemiological coupling \citep{sanchez2021mgdrive}. Numerical studies show PN implementations can reproduce SIR-type ODE dynamics at small discretization error \citep{reckell2025numerical}. Nonetheless, systematic side-by-side comparisons of ODE and PN under the same epidemiological scenarios appear relatively limited in the literature.

To address these gaps, this study provides a comprehensive framework for CHIKV outbreak analysis that bridges methodological approaches and enhances practical applicability. The framework is demonstrated through the analysis of the 2025 Foshan outbreak, providing insights into transmission dynamics and intervention effectiveness.

Overall, taking the recent CHIKV outbreak in Foshan as a case study, this paper concurrently develops comparable ODE and Petri Net implementations on a common SEICR design, grounded in the empirically informed human-side natural history of CHIKV. The framework provides empirical insights into transmission dynamics and intervention effectiveness, and mitigates the inaccuracies and stability issues that ODEs may face when modeling small, discrete, event-driven processes via time discretization.

This study contributes: first, an SEICR model is implemented in both ODE and PN, enforcing a one-to-one mapping between states and flows to ensure dynamical equivalence; second, a three-phase intervention fitting protocol for \(\beta(t)\) is tailored to align with the 2025 Foshan timeline and the three management measures enacted by local health authorities; third, dual-path \(R\) estimation with uncertainty quantification (definition-based and next-generation matrix) is designed, with comparative assessments of accuracy, computational efficiency, and stability; finally, multiple CHIKV reproduction number estimates across models and methods are reported to serve as references for future outbreaks.

\section{Materials and Methods}
\label{sec:methods}

\subsection{Data and extraction}
This study uses the official 2025 CHIKV outbreak bar charts released by the Foshan health authorities as the data source \citep{zhang2025foshan}. We digitized the daily counts via an automated image-processing pipeline, adapted from prior work. Specifically, we first load the original image with OpenCV and convert it to the HSV color space to improve channel separability. We then define color thresholds for the blue and red bars to generate binary masks. For blue, we use \([100,50,50]\)–\([130,255,255]\); for red, we take the union of \([0,50,50]\)–\([10,255,255]\) and \([170,50,50]\)–\([180,255,255]\), yielding stable color detection.

\begin{figure}[htbp]
\centering
\includegraphics[width=12cm]{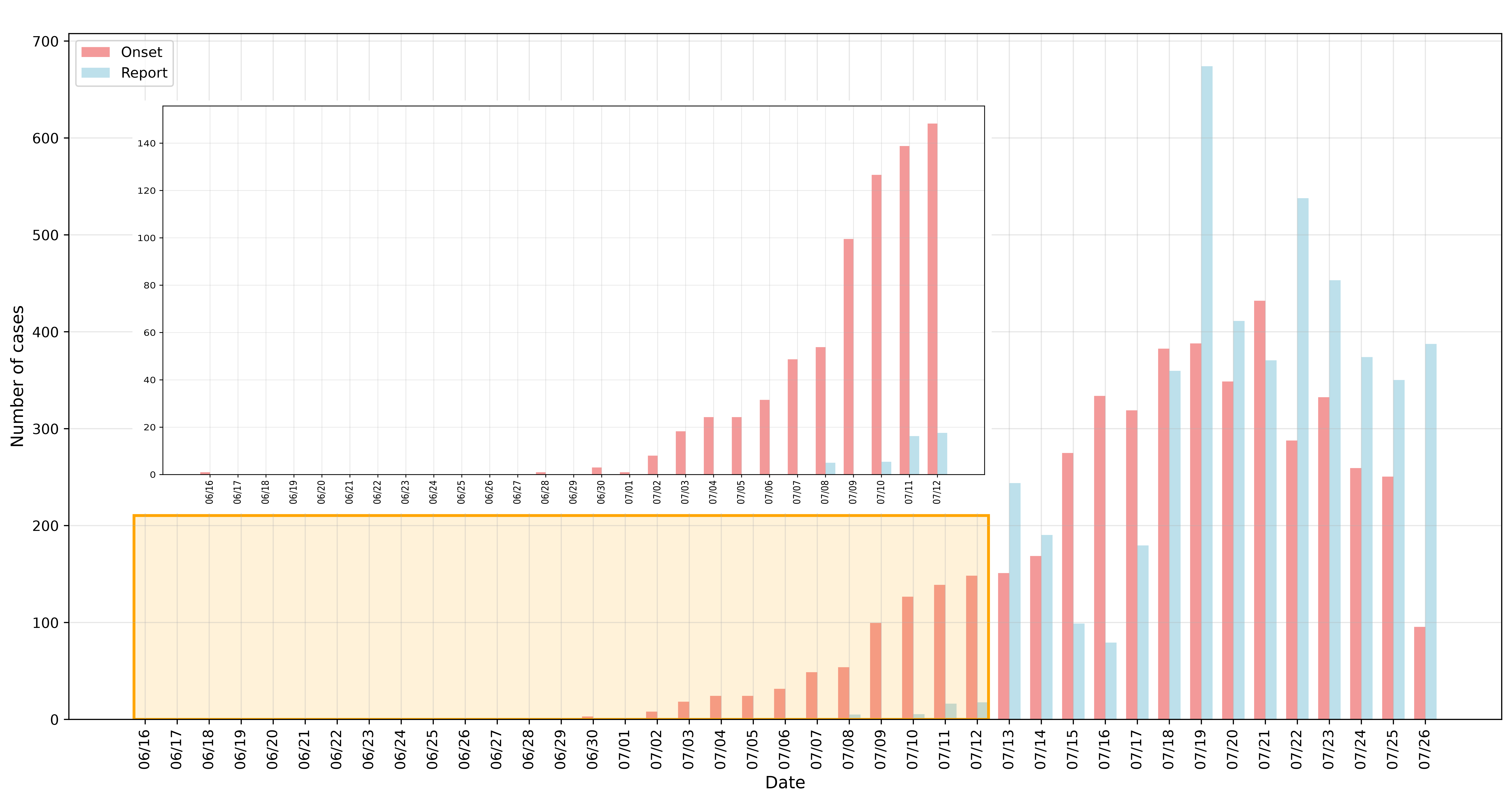}
\caption{Daily case bar plot (Foshan, 2025). See also the early-phase variant.}
\label{fig:barplots}
\end{figure}

On the mask images, we extract external contours and compute the minimum bounding rectangle for each bar to obtain the \(x\) and \(y\) coordinates, width, height, centroid, and pixel area. To reduce noise, objects with width or height smaller than 5 pixels are discarded. All bars are sorted by the \(x\)-coordinate in ascending order to match the left-to-right timeline. The geometric attributes of blue and red bars are exported separately to CSV files. To align with the official reporting scale, we calibrate bar heights using a pixel-to-count ratio: based on reported key facts (as the figure \ref{fig:barplots} shows, the incidence-by-onset-date series peaks at 432 cases on July 21; the series by report date peaks at 674 cases on July 19\citep{zhang2025foshan}), we determine the pixel-to-case conversion coefficient, and cross-validate it using the first onset record on June 16, obtaining consistent results. The resulting quality-controlled daily incidence time series is then used for model fitting and subsequent analyses.

% 中文原文：数据提取流程（从 main.tex 保留为注释）
% 本研究基于佛山市卫生部门公开的2025年CHIKV疫情官方柱状图\citep{zhang2025foshan}.在此处，本研究采用自动图像处理流程对逐日发病数进行数字化提取，这一方法是参考自。具体而言，我们首先使用OpenCV载入原始图像并转换至HSV色彩空间，以增强颜色通道的可分离性。随后针对蓝色与红色柱形分别设定阈值区间，生成二值mask。本研究蓝色阈值设置为([100,50,50]至[130,255,255])，红色阈值采用两个区间([0,50,50]至[10,255,255])与([170,50,50]至[180,255,255])并取并集，形成稳定的颜色检测结果。
% 在MASK图上，提取外部轮廓并计算每个柱形的最小外接矩形，获取横坐标x、纵坐标y、宽度、高度、中心坐标与像素面积等几何量；为降低噪声，宽与高小于5像素的目标将被剔除。所有柱形按x坐标升序排序以对应从左至右的时间顺序，并将蓝、红两类柱形的几何参数分别导出为CSV文件。为与官方统计口径一致，本研究基于像素—数值比例对柱高进行标定：参照文献报道的关键信息（按发病日期统计的病例数于7月21日达峰值432例；按报告日期统计的病例数于7月19日达峰值674例），据此确定像素与病例数的比例系数，并以6月16日首例发病记录进行交叉验证，结果一致。最终得到质量控制后的逐日发病时间序列，供后续模型拟合与分析使用。

\subsection{Model structure design and symbol explanation}
We construct a SEICR model to simulate the transmission dynamics of chikungunya fever, with the model structure selection based on the epidemiological characteristics of chikungunya\citep{who2025chik,cdc2026chik}. Chikungunya virus infection exhibits distinct clinical phases: incubation period (typically 3-7 days, range 1-12 days)\citep{cdc2026chik}, acute phase (symptomatic with high transmissibility), asymptomatic infection (approximately 3-28\% of infected individuals)\citep{who2025chik}, and chronic phase\citep{santiago2025chronic,paixao2018chronic}. These characteristics necessitate the model to distinguish between different infection states, where asymptomatic individuals, despite having lower viral loads, still possess transmission capability and contribute significantly to epidemic spread; symptomatic individuals exhibit obvious symptoms with strong transmission capacity and represent the primary targets for epidemic surveillance and control; chronic patients, while having reduced transmissibility, experience prolonged disease duration affecting disease burden assessment\citep{paixao2018chronic}.

Therefore, we provide the following symbol explanations:

\begin{table}[htbp]
\setlength{\tabcolsep}{3pt}
\caption{Symbol explanation}
\label{tab:compartments}
\centering
\small
\begin{tabularx}{\textwidth}{@{} >{\centering\arraybackslash}p{1.5cm} >{\centering\arraybackslash}p{2.5cm} >{\centering\arraybackslash}X @{} }
\toprule
\textbf{Symbol} & \textbf{English Name} & \textbf{Definition and Description} \\
\midrule
$S$ & Susceptible & Susceptible individuals: population not infected with chikungunya virus and susceptible to the disease \\
$E$ & Exposed & Exposed individuals: those infected with the virus but not yet symptomatic and non-infectious \\
$I_s$ & Symptomatic Infectious & Symptomatic infectious individuals: those with obvious clinical symptoms and infectious \\
$I_a$ & Asymptomatic Infectious & Asymptomatic infectious individuals: those infected with the virus but without obvious symptoms yet still infectious \\
$C$ & Chronic & Chronic patients: those experiencing persistent joint pain and other chronic symptoms after the acute phase \\
$R$ & Recovered & Recovered individuals: those who have completely recovered and acquired immunity \\
\bottomrule
\end{tabularx}
\end{table}

\begin{table}[htbp]
\setlength{\tabcolsep}{3pt}
\caption{Model state transition relationships and parameter explanations}
\label{tab:transitions}
\centering
\small
\begin{tabularx}{\textwidth}{@{} p{2.2cm} p{2.8cm} >{\raggedright\arraybackslash}X @{} }
\toprule
\textbf{Transition Process} & \textbf{Mathematical Expression} & \textbf{Biological Significance and Parameter Explanation} \\
\midrule
Infection process & $S \xrightarrow{\lambda} E$ & Susceptible individuals enter the exposed state through contact with infected individuals, where $\lambda = \beta(t) (I_s + \epsilon I_a)/N$ is the force of infection \\
Exposed transition & $E \xrightarrow{p\sigma} I_s$ & Exposed individuals transition to symptomatic infectious state with probability $p$, where $\sigma$ is the exposed transition rate \\
 & $E \xrightarrow{(1-p)\sigma} I_a$ & Exposed individuals transition to asymptomatic infectious state with probability $(1-p)$ \\
Acute phase transition & $I_s \xrightarrow{q\gamma} C$ & Symptomatic infectious individuals progress to chronic state with probability $q$, where $\gamma$ is the acute recovery rate \\
 & $I_s \xrightarrow{(1-q)\gamma} R$ & Symptomatic infectious individuals recover directly with probability $(1-q)$ \\
 & $I_a \xrightarrow{\gamma} R$ & Asymptomatic infectious individuals recover directly without entering the chronic phase \\
Chronic phase transition & $C \xrightarrow{\omega} R$ & Chronic patients eventually recover, where $\omega$ is the chronic recovery rate \\
Demographic dynamics & $\Lambda, \mu$ & All compartments consider the effects of population birth\&immigration ($\Lambda$) and natural death ($\mu$) \\
\bottomrule
\end{tabularx}
\end{table}

\subsection{Model Parameter}
This section summarizes the key parameters, symbols, values, units, along with brief explanations of their physical/epidemiological meanings. $N$ represents the resident population size of the study area (unit: persons), with an estimated total population of 9,698,900 in Foshan City at the end of 2024. $\\Lambda$ denotes the population inflow rate (birth\&immigration flux) measured in persons/day, while $\\mu$ represents the natural death rate (unit: day$^{-1}$). These demographic parameters are calculated based on the latest 2024 population statistics for Foshan City \citep{foshan2024statistical}. The resident population net growth in 2024 was 83,500 individuals, comprising 56,346 from natural population growth (based on a 5.86‰ natural growth rate) and 27,154 from net immigration, yielding $\\Lambda = 83,500/365 = 229$ persons/day, including 154 daily births and 74 daily net immigrants. The $\\mu$ parameter (daily death rate) is calculated from the 2024 registered population death rate of 5.04‰: $\\mu = 5.04‰/365 = 1.381 \\times 10^{-5}$ day$^{-1}$.

Regarding transmission and progression parameters, $\sigma$ represents the inverse of the incubation period (unit: day$^{-1}$); $\gamma$ is the acute recovery rate (unit: day$^{-1}$); $p$ denotes the proportion of symptomatic cases (dimensionless); among chronic-related parameters, $q$ represents the proportion progressing from acute to chronic phase (dimensionless), and $\omega$ is the chronic recovery rate (unit: day$^{-1}$). $\epsilon$ represents the relative infectiousness of asymptomatic infections (dimensionless). The calculation is based on viral load difference data from \citep{[9],[10]}: symptomatic patients had a median viral load of $5.6 \times 10^5$ pfu/mL, while asymptomatic patients had $3.4 \times 10^3$ pfu/mL (165-fold difference), along with mosquito infection probability data from \citep{[8]}. Through viral load logarithmic transformation and linear interpolation methods, the estimated mosquito infection probabilities are 84.6\% for symptomatic patients and 31.7\% for asymptomatic patients, yielding $\epsilon = 31.7\%/84.6\% = 0.375$.In the model, epsilon is used for transmission intensity calculation $\lambda = \beta(t) (I_s + \epsilon I_a)/N$ and effective infectious period calculation.

The unit choices maintain consistency with common compartmental model notation: rate parameters are measured in day$^{-1}$; proportion parameters are dimensionless; population/flux are measured in "persons" and "persons/day", respectively.

\begin{table}[htbp]
\setlength{\tabcolsep}{3pt}
  \caption{Model parameters}
  \label{tab:params}
\centering
\small
\begin{tabularx}{\textwidth}{@{} >{\centering\arraybackslash}p{2.2cm} >{\centering\arraybackslash}p{1.5cm} >{\centering\arraybackslash}p{1.8cm} >{\centering\arraybackslash}p{1.5cm} >{\centering\arraybackslash}X @{} }
  \toprule
  \textbf{Name} & \textbf{Symbol} & \textbf{Value} & \textbf{Unit} & \textbf{Source}\\
  \midrule
  Total population & $N$ & 9,698,900 & persons & \citep{foshan2024statistical} \\
  birth\&immigration rate & $\Lambda$ & 229 & persons/day & \citep{zhang2025foshan} \\
  Natural death rate & $\mu$ & $1.381\times10^{-5}$ & day$^{-1}$ & \citep{zhang2025foshan,foshan2024statistical,foshan2023statistical} \\
  Incubation inverse & $\sigma$ & 0.0714 & day$^{-1}$ & \citep{zhang2025foshan} \\
  Symptomatic proportion & $p$ & 0.75 & -- & \citep{[3],[6]} \\
  Recovery rate & $\gamma$ & 0.143 & day$^{-1}$ & \citep{[3],[4]} \\
  Chronic progression & $q$ & 0.43 & -- & \citep{[8]} \\
  Chronic recovery & $\omega$ & 0.0026 & day$^{-1}$ & \citep{[8]} \\
  Asymptomatic rel. inf. & $\epsilon$ & 0.375 & -- & \citep{[9],[10]} \\
  \bottomrule
  \end{tabularx}
  \end{table}

\subsection{Compartmental ODE model}
We adopt a six-compartment SEICR model with state $(S,E,I_s,I_a,C,R)$. The dynamics follow the balance equations described in our documentation, with demographic inflow/outflow and transition rates $\sigma,\gamma,\omega,\mu, p, q$. The solver and parameter registry are implemented within our analysis codebase.

Firstly, we give the force of infection is defined as:
\begin{equation}
\lambda = \frac{\beta(t) (I_s + \epsilon I_a)}{N} \label{eq:force_infection}
\end{equation}

Thus, the system of ordinary differential equations governing the SEICR model dynamics is given by:

\begin{align}
\frac{dS}{dt} &= \Lambda - \lambda S - \mu S \label{eq:dS} \\
\frac{dE}{dt} &= \lambda S - \sigma E - \mu E \label{eq:dE} \\
\frac{dI_s}{dt} &= p\sigma E - \gamma I_s - \mu I_s \label{eq:dIs} \\
\frac{dI_a}{dt} &= (1-p)\sigma E - \gamma I_a - \mu I_a \label{eq:dIa} \\
\frac{dC}{dt} &= q\gamma I_s - \omega C - \mu C \label{eq:dC} \\
\frac{dR}{dt} &= (1-q)\gamma I_s + \gamma I_a + \omega C - \mu R \label{eq:dR}
\end{align}

and the total population satisfies:
\begin{equation}
N = S + E + I_s + I_a + C + R \label{eq:total_pop}
\end{equation}

The model parameters are defined as follows: $\Lambda$ is the population birth\&immigration rate, $\mu$ is the natural death rate, $\beta(t)$ is the time-dependent transmission rate, $\epsilon$ is the relative infectiousness of asymptomatic infections, $\sigma$ is the exposed transition rate, $p$ is the proportion of exposed individuals who become symptomatic, $\gamma$ is the acute recovery rate, $q$ is the proportion of symptomatic individuals who progress to chronic phase, and $\omega$ is the chronic recovery rate.

\subsection{Petri Net model construction}
Based on the aforementioned advantages of Petri Nets in epidemiological modeling, we construct a Petri Net implementation comparable to the SEICR ODE model. This PN model contains places corresponding to SEICR compartments and transitions representing infection, progression, recovery, mortality, and birth processes. Variable-rate transitions implement the same rate functions as the ODE terms, ensuring structural similarity.

The Petri Net model can be formally defined as a six-tuple 
$PN = (P, T, F, W, M_0, \Lambda)$, with symbol explanations as follows:

\begin{table}[htbp]
\setlength{\tabcolsep}{3pt}
\caption{Petri Net model symbol explanations}
\label{tab:pn_symbols}
\centering
\small
\begin{tabularx}{\textwidth}{@{} >{\centering\arraybackslash}p{1.5cm} >{\centering\arraybackslash}p{5.0cm} >{\centering\arraybackslash}X @{} }
\toprule
\textbf{Symbol} & \textbf{Definition} & \textbf{Description} \\
\midrule
$P$ & $\{S, E, I_s, I_a, C, R\}$ & Set of places, corresponding to the six compartments of the SEICR model \\
 & $\{t_{inf}, t_{exp\_s}, t_{exp\_a},$ &  \\
$T$ & $t_{rec\_s}, t_{rec\_a}, t_{chr}, t_{rec\_c},$ & Set of transitions, representing all state transition processes \\
 & $t_{birth}, t_{death}\}$ & \\
$F$ & $\subseteq (P \times T) \cup (T \times P)$ & Flow relation, defining connections between places and transitions \\
$W$ & $F \rightarrow \mathbb{R}^+$ & Weight function, defining arc weights \\
$M_0$ & $P \rightarrow \mathbb{N}$ & Initial marking, representing initial population in each compartment \\
$\Lambda$ & $T \rightarrow \mathbb{R}^+$ & Transition rate function, defining firing rates for each transition \\
\bottomrule
\end{tabularx}
\end{table}

The rate functions for each transition remain consistent with the corresponding terms in the ODE model:
\begin{table}[htbp]
\setlength{\tabcolsep}{3pt}
\caption{Petri Net model transition rate functions}
\label{tab:pn_rates}
\centering
\small
\begin{tabularx}{\textwidth}{@{} >{\centering\arraybackslash}p{2.0cm} >{\centering\arraybackslash}p{4.0cm} >{\centering\arraybackslash}X @{} }
\toprule
\textbf{Transition} & \textbf{Rate Function} & \textbf{Description} \\
\midrule
$t_{inf}$ & $\lambda_{inf}(t) = \frac{\beta(t) (I_s(t) + \epsilon I_a(t))}{N(t)}$ & Infection transition \\
$t_{exp\_s}$ & $\lambda_{exp\_s}(t) = p\sigma E(t)$ & Exposed to symptomatic infectious transition \\
$t_{exp\_a}$ & $\lambda_{exp\_a}(t) = (1-p)\sigma E(t)$ & Exposed to asymptomatic infectious transition \\
$t_{rec\_s}$ & $\lambda_{rec\_s}(t) = (1-q)\gamma I_s(t)$ & Symptomatic infectious direct recovery transition \\
$t_{chr}$ & $\lambda_{chr}(t) = q\gamma I_s(t)$ & Symptomatic infectious to chronic transition \\
$t_{rec\_a}$ & $\lambda_{rec\_a}(t) = \gamma I_a(t)$ & Asymptomatic infectious recovery transition \\
$t_{rec\_c}$ & $\lambda_{rec\_c}(t) = \omega C(t)$ & Chronic patient recovery transition \\
$t_{birth}$ & $\lambda_{birth}(t) = \Lambda$ & Population birth transition \\
$t_{death}$ & $\lambda_{death}(t) = \mu N(t)$ & Natural death transition \\
\bottomrule
\end{tabularx}
\end{table}
\begin{align}
\lambda_{inf}(t) &= \frac{\beta(t) (I_s(t) + \epsilon I_a(t))}{N(t)} \label{eq:pn_inf} \\
\lambda_{exp\_s}(t) &= p\sigma E(t) \label{eq:pn_exp_s} \\
\lambda_{exp\_a}(t) &= (1-p)\sigma E(t) \label{eq:pn_exp_a} \\
\lambda_{rec\_s}(t) &= (1-q)\gamma I_s(t) \label{eq:pn_rec_s} \\
\lambda_{chr}(t) &= q\gamma I_s(t) \label{eq:pn_chr} \\
\lambda_{rec\_a}(t) &= \gamma I_a(t) \label{eq:pn_rec_a} \\
\lambda_{rec\_c}(t) &= \omega C(t) \label{eq:pn_rec_c} \\
\lambda_{birth}(t) &= \Lambda \label{eq:pn_birth} \\
\lambda_{death}(t) &= \mu N(t) \label{eq:pn_death}
\end{align}

\subsection{Three-phase intervention modeling}
Aligned with the actual intervention timeline in Foshan, we model the transmission rate $\beta(t)$ as a piecewise-constant function over three intervention phases. The concrete measures included the launch of nucleic acid testing in medical institutions on July 14, followed by a three-day intensive campaign for the clearance of mosquito breeding sites from July 18 to July 20 \citep{zhang2025foshan}.

Based on the official timeline and the implemented control measures, we partition the outbreak into four phases:

\begin{table}[htbp]
\setlength{\tabcolsep}{3pt}
\caption{Intervention phase definitions for the Foshan chikungunya outbreak}
\label{tab:intervention_phases}
\centering
\small
\begin{tabularx}{\textwidth}{@{} >{\centering\arraybackslash}p{1.8cm} >{\centering\arraybackslash}p{2.8cm} >{\centering\arraybackslash}p{1.8cm} >{\centering\arraybackslash}X @{} }
\toprule
\textbf{Phase} & \textbf{Date range} & \textbf{Transmission rate} & \textbf{Primary interventions} \\
\midrule
Initial & 2025-06-16 to 2025-07-09 & $\beta_0$ & No control; natural transmission \\
Phase 1 & 2025-07-09 to 2025-07-14 & $\beta_1$ & Early surveillance and case detection \\
Phase 2 & 2025-07-14 to 2025-07-18 & $\beta_2$ & Nucleic acid testing launched in medical institutions \\
Phase 3 & After 2025-07-18 & $\beta_3$ & Three-day intensive clearance of mosquito breeding sites\\
\bottomrule
\end{tabularx}
\end{table}

\subsection{Parameter fitting method}
To ensure fidelity to the actual implementation, we adopt an isomorphic three-phase fitting workflow for both the ODE and Petri Net pathways. Anchored to the official Foshan outbreak timeline, we set the outbreak start date to 2025-06-16 and define three intervention onset dates—July 9 (Phase 1), July 14 (Phase 2), and July 18 (Phase 3)—denoted by $\mathrm{Reff\_1}$, $\mathrm{Reff\_2}$, and $\mathrm{Reff\_3}$, respectively. These dates are converted to days since start, $\{t_1,t_2,t_3\}$, for piecewise simulation. Observations are the daily incidence by onset date \citep{zhang2025foshan}.

Both implementations (ODE and PN) adopt a piecewise-constant transmission rate and perform continuous simulations over four contiguous segments, using the terminal state of one segment as the initial condition of the next. Let the phase boundaries satisfy $0<t_1<t_2<t_3$; then
\begin{equation}
\beta(t)=\begin{cases}
\beta_0, & 0\le t\le t_1,\\
\beta_1, & t_1<t\le t_2,\\
\beta_2, & t_2<t\le t_3,\\
\beta_3, & t>t_3~.
\end{cases}
\label{eq:piecewise_beta}
\end{equation}
At any time $t$, the daily incidence flow is extracted from the exposed compartment as
\begin{equation}
\operatorname{inc}(t)=p\,\sigma\,E(t).\label{eq:incidence}
\end{equation}
Let the calendar-day index set be $\{d_j\}_{j=1}^{D}$. We apply the linear interpolation operator $\mathcal{I}_{\mathrm{lin}}$ to map \eqref{eq:incidence} onto the observation grid, yielding predictions $\hat y_j=\mathcal{I}_{\mathrm{lin}}\big(\operatorname{inc}\big)(d_j)$\citep{press2007numerical}.

We then minimize the root-mean-squared error (RMSE):
\begin{equation}
\mathrm{RMSE}(\beta_0,\beta_1,\beta_2,\beta_3)=\sqrt{\frac{1}{D}\sum_{j=1}^{D}\big(\hat y_j-y_j\big)^2}~,
\label{eq:rmse}
\end{equation}
where $y_j$ denotes the observed daily incidence on day $d_j$. We impose the following constraints:
\begin{equation}
\beta_i\in[0.01,15.0]~\text{day}^{-1}\ (i=0,1,2,3),\qquad \beta_0>\beta_1\ge\beta_2>\beta_3>0~.
\label{eq:constraints}
\end{equation}

Subject to \eqref{eq:constraints} and the box bounds, we employ Differential Evolution (DE)\citep{storn1997de} to globally minimize \eqref{eq:rmse}.
%  The implementation settings are $\texttt{seed}=42$, $\texttt{maxiter}=400$, $\texttt{popsize}=100$, and $\texttt{tol}=10^{-12}$. In both pathways, the initial number of infected individuals is fixed at 1\citep{zhang2025foshan}.

\subsection{Reproduction number $R_0$ estimation}
This study estimates the reproduction number $R$ across all phases using two approaches: the "definition method" and the "next-generation matrix (NGM) method". According to the definition method \citep{guo2022computing}, only $I_s$ (symptomatic infectious individuals) and $I_a$ (asymptomatic infectious individuals) contribute to subsequent transmission in the model. The expected number of secondary cases generated by a typical initial individual can be weighted by pathway proportions and relative infectiousness. When natural mortality $\mu$ is present, the definition method yields:
\begin{equation}
R_{\mathrm{def}}(\beta)
= \beta\,\frac{\sigma}{\sigma+\mu}\left( \frac{p}{\gamma+\mu}
 + \frac{(1-p)\,\epsilon}{\gamma+\mu} \right),
\label{eq:R_def_full}
\end{equation}
where the factor $\sigma/(\sigma+\mu)$ reflects the survival proportion from the latent period to the infectious period, and $1/(\gamma+\mu)$ represents the average infectious period for $I_s$ and $I_a$. For short outbreak durations or when $\mu\ll \sigma,\gamma$, equation \eqref{eq:R_def_full} approximates to the commonly used expression:
\begin{equation}
R_{\mathrm{def}}(\beta) \approx \beta\left( \frac{p}{\gamma} + \frac{(1-p)\,\epsilon}{\gamma} \right).
\label{eq:R_def_simple}
\end{equation}
For phase-specific estimation, substituting $\beta\in\{\beta_0,\beta_1,\beta_2,\beta_3\}$ into equation \eqref{eq:R_def_full} (or equation \eqref{eq:R_def_simple} when $\mu$ is negligible) yields $R_{\mathrm{def},k}$ ($k=0,1,2,3$).

For the next-generation matrix (NGM) method in ODEs, we first linearize the system at the disease-free equilibrium (DFE) where $(S^*,E^*,I_s^*,I_a^*,C^*,R^*) = (N,0,0,0,0,0)$ \citep{vanden2002reproduction}. The disease-free equilibrium represents the state where no infection is present in the population, where the susceptible population equals the total population and all infectious compartments are empty. This equilibrium serves as the reference point for linearization and reproduction number calculation. Let the infectious subsystem be $(E,I_s,I_a)$ and define the new infection matrix and transition matrix as:
\begin{equation}
F = \begin{pmatrix}
0 & \beta & \beta\,\epsilon\\
0 & 0 & 0\\
0 & 0 & 0
\end{pmatrix},\qquad
V = \begin{pmatrix}
\sigma+\mu & 0 & 0\\
-p\,\sigma & \gamma+\mu & 0\\
-(1-p)\,\sigma & 0 & \gamma+\mu
\end{pmatrix}.
\label{eq:FV}
\end{equation}
The next-generation matrix is then $K=FV^{-1}$, and the reproduction number is its spectral radius $R_{\mathrm{NGM}}=\rho(K)$ \citep{vanden2002reproduction}. For phase $k$, replacing $\beta$ in $F$ with $\beta_k$ yields $R_{\mathrm{NGM},k}$.

In the Petri Net framework, the construction of the next-generation matrix method is based on the same infectious subsystem $(E,I_s,I_a)$. This study references Segovia's geometric interpretation method \citep{segovia2025petri} and the work of Reckell et al. \citep{reckell2025numerical}, first identifying the Kermack-McKendrick modules in the Petri Net: the susceptible module ($S$), infection-process module ($E$), and infectious module ($I_s,I_a$).

Similarly, the disease-free equilibrium (DFE) in the Petri Net framework corresponds to the marking vector $m^*=(N,0,0,0,0,0)$, representing that the susceptible place contains the entire population while all infection-related places are empty. This equilibrium serves as the reference state for Petri Net linearization and reproduction number calculation \citep{vanden2002reproduction}.

For the infectious subsystem $(E,I_s,I_a)$, the Petri Net's Jacobian matrix $J$ is decomposed at the DFE as $J=F-V$, where:
\begin{equation}
F^{\mathrm{PN}} = \begin{pmatrix}
0 & \beta & \beta\,\epsilon\\
0 & 0 & 0\\
0 & 0 & 0
\end{pmatrix},\qquad
V^{\mathrm{PN}} = \begin{pmatrix}
\sigma+\mu & 0 & 0\\
-p\,\sigma & \gamma+\mu & 0\\
-(1-p)\,\sigma & 0 & \gamma+\mu
\end{pmatrix}.
\label{eq:FV_PN}
\end{equation}

The construction of matrix $F^{\mathrm{PN}}$ is based on the rate function of the new infection transition in the Petri Net: the infection transition $t_{\mathrm{inf}}$ from susceptible $S$ to exposed $E$, with rate function $\lambda_{\mathrm{inf}}(t) = \beta(t) S(t) (I_s(t) + \epsilon I_a(t))/N(t)$. At the DFE where $S \approx N$, the non-zero elements of $F^{\mathrm{PN}}$ correspond to the partial derivatives of the infection transition with respect to $I_s$ and $I_a$.

Matrix $V^{\mathrm{PN}}$ contains all non-infection transitions: transitions from $E$ to $I_s$ (probability $p$, rate $\sigma$), transitions from $E$ to $I_a$ (probability $1-p$, rate $\sigma$), recovery transitions from $I_s$ and $I_a$ (rate $\gamma$), and natural death (rate $\mu$).

Therefore, the next-generation matrix in the Petri Net framework is $K^{\mathrm{PN}} = (F^{\mathrm{PN}})(V^{\mathrm{PN}})^{-1}$, and the basic reproduction number is $R_{\mathrm{NGM}}^{\mathrm{PN}} = \rho(K^{\mathrm{PN}})$.

For each phase $k \in \{0,1,2,3\}$, replacing $\beta$ in $F^{\mathrm{PN}}$ with $\beta_k$ yields the phase-specific reproduction number $R_{\mathrm{NGM},k}^{\mathrm{PN}}$.

\subsection{Confidence interval estimation methods}

For confidence interval estimation, this study employs a Monte Carlo sampling approach to calculate confidence intervals for the basic reproduction number. This method evaluates the statistical properties of model outputs through parameter uncertainty propagation. Let the model parameter vector be $\boldsymbol{\theta} = (\beta, p, \gamma, \epsilon, \sigma, q, \omega, \mu)^T$, where $\beta$ is the transmission coefficient and other parameters are fixed values. Parameter uncertainty is modeled as follows:

\begin{equation}
\beta^{(i)} \sim \mathcal{N}(\beta, \sigma_{\beta}^2), \quad i = 1, 2, \ldots, N
\end{equation}

where $\sigma_{\beta} = 0.05 \times \beta$ is the standard deviation of the transmission coefficient, and $N = 1000$ is the number of Monte Carlo samples. For each sampled parameter $\beta^{(i)}$, the corresponding basic reproduction number is calculated:

\begin{equation}
R_{\mathrm{def}}^{(i)} = \beta^{(i)} \times \left(\frac{p}{\gamma} + \frac{(1-p)\epsilon}{\gamma}\right)
\end{equation}

\begin{equation}
R_{\mathrm{NGM}}^{(i)} = \rho\left(F(\beta^{(i)}) V^{-1}\right)
\end{equation}

Subsequently, let $\{R^{(1)}, R^{(2)}, \ldots, R^{(N)}\}$ be the $N$ Monte Carlo samples, with confidence level $\alpha$ (set to $\alpha = 0.95$ in this study), then the $100(1-\alpha)\%$ confidence interval is:

\begin{equation}
CI_{1-\alpha} = \left[R_{\alpha/2}, R_{1-\alpha/2}\right]
\end{equation}

where $R_{\alpha/2}$ and $R_{1-\alpha/2}$ are the $\alpha/2$ and $1-\alpha/2$ percentiles of the samples, respectively:

\begin{equation}
R_{\alpha/2} = \text{percentile}(\{R^{(i)}\}, 100 \times \alpha/2)
\end{equation}

\begin{equation}
R_{1-\alpha/2} = \text{percentile}(\{R^{(i)}\}, 100 \times (1-\alpha/2))
\end{equation}

This method quantifies the uncertainty in model outputs through parameter perturbation, providing statistical reliability assessment for epidemiological parameter estimation.

%%%%%%%%%%%%%%%%%%%%%%%%%%%%%%%%%%%%%%%%%%
\section{Results}
\label{sec:results}

\subsection{Three-phase intervention fitting results}
\begin{figure}[htbp]
\centering
\includegraphics[width=14cm]{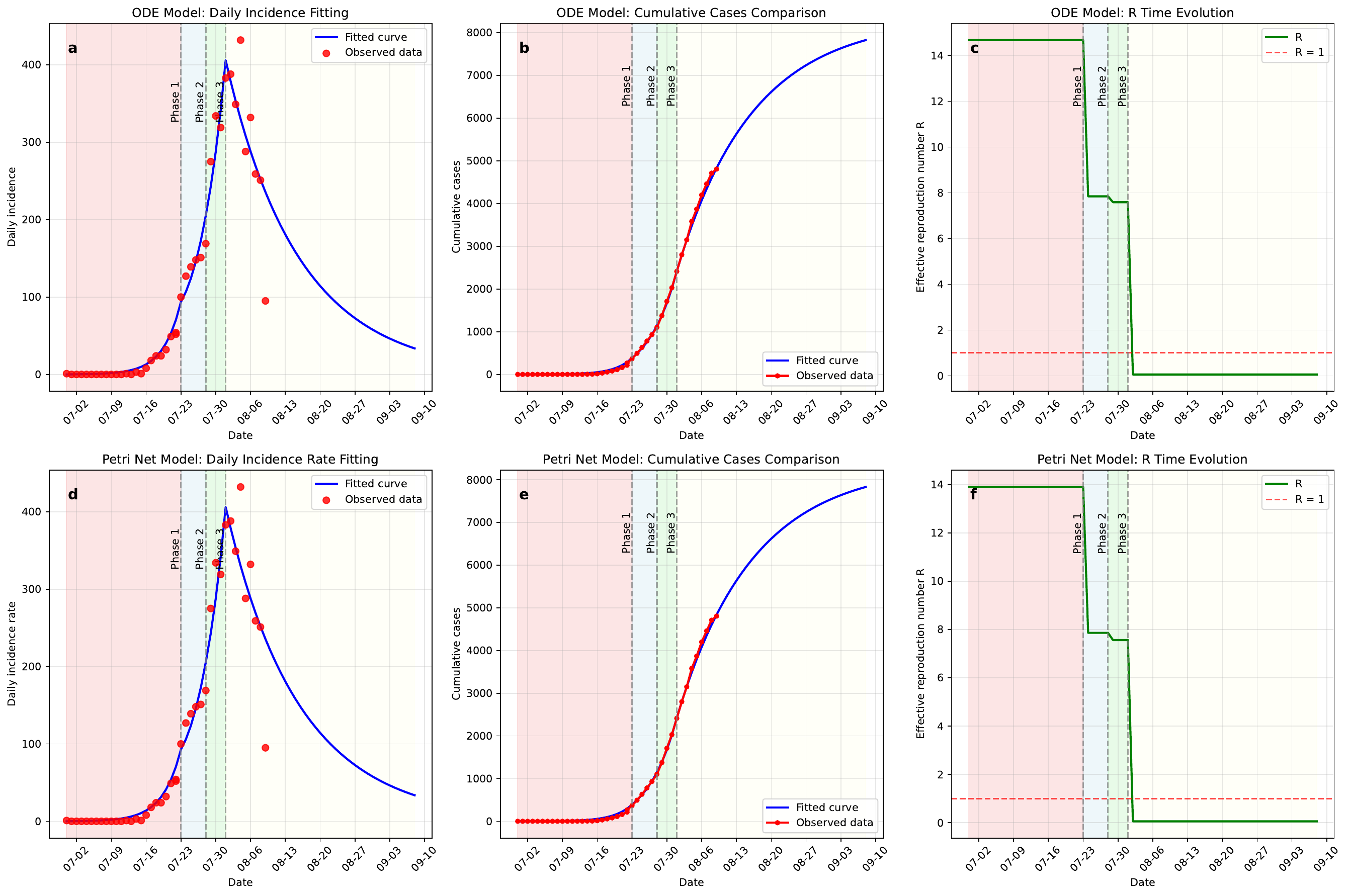}
\caption{Three-phase intervention fitting (ODE model vs Petri Net model comparison). The figure displays the fitting and prediction results of both models across three intervention phases for daily new cases, cumulative cases, and effective reproduction number (R). Vertical dashed lines indicate the start dates of the three intervention phases: Phase 1 (July 9-14), Phase 2 (July 14-18), and Phase 3 (After July 18). Subplots: (a) ODE model daily incidence fitting, (b) ODE model cumulative cases comparison, (c) ODE model effective reproduction number evolution, (d) Petri Net model daily incidence fitting, (e) Petri Net model cumulative cases comparison, (f) Petri Net model effective reproduction number evolution.}
\label{fig:fit}
\end{figure}

Figure \ref{fig:fit} presents the three-phase intervention fitting results for CHIKV outbreak data using both ODE (top row) and Petri Net (bottom row) models. The left column shows the fitting of daily new cases, the middle column compares cumulative cases, and the right column displays the temporal evolution of the effective reproduction number (R). To ensure comparability between different models, all R values are calculated using the next-generation matrix method.

The fitting results demonstrate that both models' curves effectively track the overall trend of observed data. The ODE model achieves a mean absolute error (MAE) of 18.77 cases, root mean square error (RMSE) of 36.52 cases, and mean absolute percentage error (MAPE) of 68.3\%. The Petri Net model shows MAE of 18.91 cases, RMSE of 36.54 cases, and MAPE of 73.6\%.

The observed data indicates that the epidemic peaked on day 35 with 432 daily new cases. The ODE model predicts the peak on day 32 with 406.0 cases, while the Petri Net model predicts day 32 with 406.1 cases. The predicted peak time is 3 days earlier than observed, with peak value errors of 6.0\% and 6.0\%, respectively. According to model predictions, the epidemic will continue to decline after the peak, with daily new cases expected to drop below 10 cases around day 45, below 1 case around day 55, and achieve basic control (daily new cases $<0.1$) around day 65.

The evolution of the effective reproduction number ($R$) clearly reflects the effectiveness of the three-phase intervention measures. In the initial phase, the ODE model estimates $R_0 = 14.67$ and the Petri Net model estimates $R_0 = 13.90$, showing reasonable agreement between the two models. As intervention measures are implemented, the $R$ values decrease significantly: after Phase 1 intervention, the ODE model shows $R_{\mathrm{eff}} = 7.85$ and the Petri Net model shows $R_{\mathrm{eff}} = 7.86$; after Phase 2 intervention, the ODE model shows $R_{\mathrm{eff}} = 7.59$ and the Petri Net model shows $R_{\mathrm{eff}} = 7.56$; after Phase 3 intervention, both models' $R_{\mathrm{eff}}$ values drop to 0.059, well below the transmission threshold of 1.0.

The changes in transmission coefficients reflect the effectiveness of intervention measures. The ODE model shows that Phase 1 intervention reduces the transmission coefficient from $\beta_0 = 2.49$ to $\beta_1 = 1.33$ (47\% reduction), Phase 2 further reduces it to $\beta_2 = 1.29$ (48\% reduction), and Phase 3 dramatically reduces it to $\beta_3 = 0.010$ (99\% reduction). The Petri Net model shows similar results: $\beta_0 = 2.36 \rightarrow \beta_1 = 1.33$ (43\% reduction) $\rightarrow \beta_2 = 1.28$ (46\% reduction) $\rightarrow \beta_3 = 0.010$ (99\% reduction).

Comparing the two models, both ODE and Petri Net models demonstrate high consistency in fitting this CHIKV outbreak data. Their total predicted case numbers (ODE: 7825 cases, Petri Net: 7831 cases) are very close, and their R value estimates show close agreement, with maximum differences not exceeding 5.3\%. This indicates that both models have similar explanatory and predictive capabilities for this CHIKV outbreak data. However, both models' predicted total case numbers are higher than the actual observed total (4806 cases), which may be attributed to our models underestimating the effectiveness of control measures.

Table \ref{tab:phase} summarizes the phase-specific transmission coefficients and reproduction numbers derived from the fitted models.

\begin{table}[htbp]
\setlength{\tabcolsep}{3pt}
\caption{Three-phase fitted parameters and reproduction numbers}
\label{tab:phase}
\centering
\small
\begin{tabularx}{\textwidth}{@{} >{\centering\arraybackslash}p{2.0cm} >{\centering\arraybackslash}p{1.5cm} >{\centering\arraybackslash}p{2.0cm} >{\centering\arraybackslash}p{2.0cm} >{\centering\arraybackslash}p{2.0cm} @{} }
\toprule
\textbf{Parameter} & \textbf{Phase} & \textbf{ODE Model} & \textbf{Petri Net Model} & \textbf{Relative Difference (\%)}\\
\midrule
$\beta_0$ & Initial & 2.49 day$^{-1}$ & 2.36 day$^{-1}$ & 5.26\\
$\beta_1$ & Phase 1 & 1.33 day$^{-1}$ & 1.33 day$^{-1}$ & 0.22\\
$\beta_2$ & Phase 2 & 1.29 day$^{-1}$ & 1.28 day$^{-1}$ & 0.32\\
$\beta_3$ & Phase 3 & 0.010 day$^{-1}$ & 0.010 day$^{-1}$ & $\approx 0$\\
$R_0$ & Initial & 14.67 & 13.90 & 5.26\\
$R_{\mathrm{eff},1}$ & Phase 1 & 7.85 & 7.86 & 0.22\\
$R_{\mathrm{eff},2}$ & Phase 2 & 7.59 & 7.56 & 0.32\\
$R_{\mathrm{eff},3}$ & Phase 3 & 0.059 & 0.059 & $\approx 0$\\
\bottomrule
\end{tabularx}
\end{table}

\subsection{Model Validation Analysis and Comparison}

\begin{figure}[htbp]
\centering
\includegraphics[width=14cm]{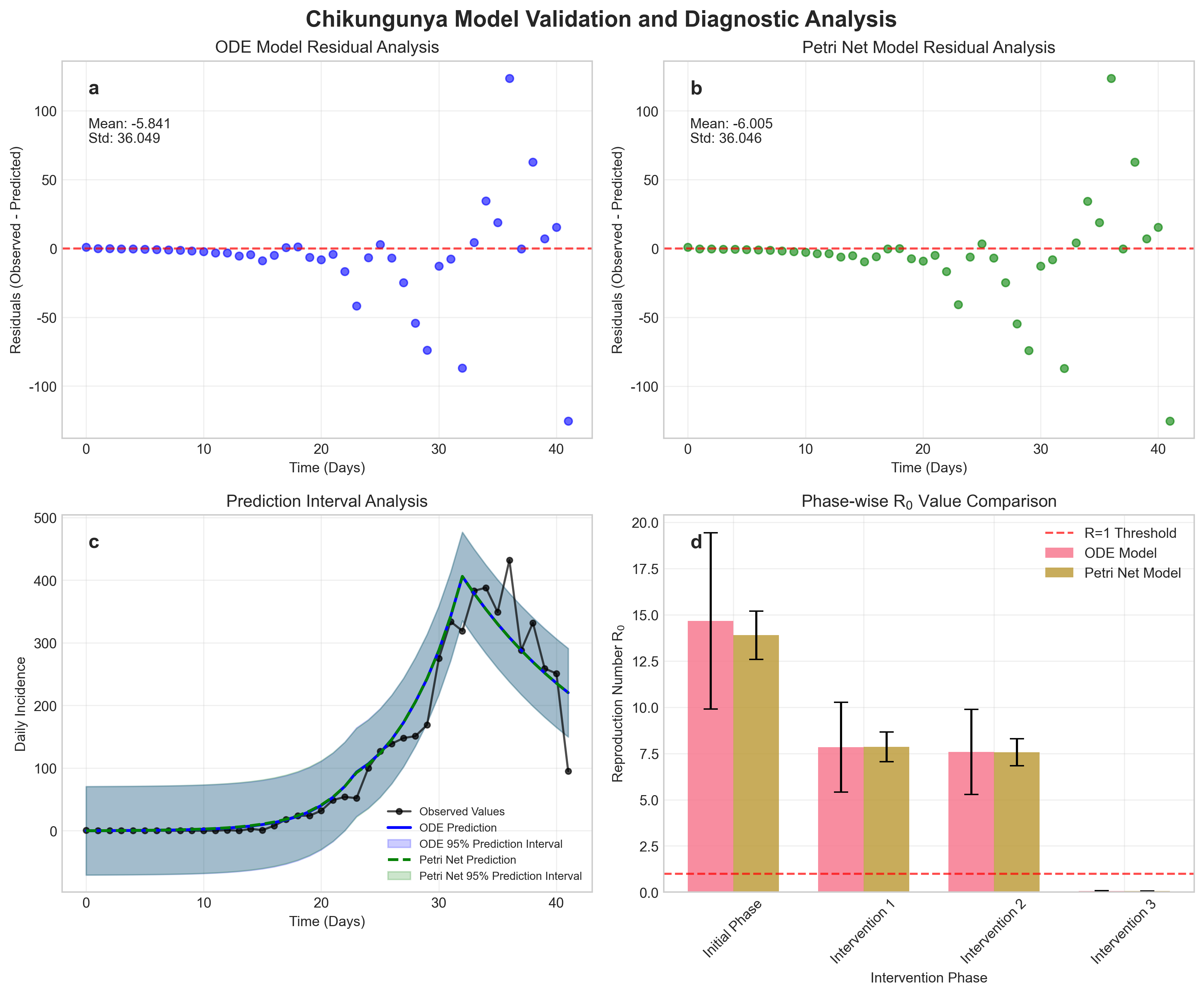}
\caption{Model validation and diagnostic analysis. This figure presents comprehensive validation results of ODE and Petri Net models, including residual analysis, prediction interval analysis, and phased $R_0$ value comparison. Subplots: (a) ODE model residual analysis, (b) Petri Net model residual analysis, (c) prediction interval analysis, (d) phased $R_0$ value comparison.}
\label{fig:model_validation}
\end{figure}

Figure \ref{fig:model_validation} presents the comprehensive validation and diagnostic analysis results of the Chikungunya epidemiological models, comparing the performance of Ordinary Differential Equation (ODE) models and Petri Net models in residual analysis, prediction intervals, and phased reproduction number ($R_0$) comparisons.

Subplots (a) and (b) display the residual distributions (observed - predicted values) for the ODE model and Petri Net model, respectively. The residual statistics for both models are as follows: the ODE model exhibits a residual mean of -5.841 with a standard deviation of 36.049, while the Petri Net model shows a residual mean of -6.005 with a standard deviation of 36.046. However, both models demonstrate similar systematic patterns in their residuals rather than the ideal random white noise characteristics. During the early epidemic phase (approximately the first 20 days), residuals remain close to zero, indicating good model fitting. However, during and around the epidemic peak period (approximately days 20-35), residuals systematically become negative, followed by large positive fluctuations. This non-random pattern suggests that the models may not fully capture the dynamic changes induced by intervention measures or other external factors, particularly during the epidemic decline phase, where model predictions exhibit systematic bias and systematic underestimation of the excess effectiveness achieved through governmental epidemic control measures.

Subplot (c) compares the prediction trajectories of both models with their 95\% prediction intervals. The vast majority of observed values fall within the similar 95\% prediction intervals of both models, validating the predictive capability of the models. Subplot (d) compares the reproduction number $R_0$ estimates across four different phases (initial phase and three intervention phases). Both models consistently reveal the phase-wise decline trend of $R_0$ values with the implementation of intervention measures. The initial phase $R_0$ value is significantly higher than 1.0, indicating rapid epidemic transmission. Subsequent intervention measures markedly reduce the $R_{\mathrm{eff}}$ values, ultimately achieving $R_{\mathrm{eff}} < 1.0$ in the third intervention phase, signifying effective epidemic transmission control.

Both models demonstrate nearly identical performance in terms of goodness-of-fit, predictive capability, and R-value inference. The RMSE is 30.98, and the average relative difference in phased transmission rates $\beta(t)$ is merely 1.45\%, with R estimates showing nearly identical results within confidence intervals (Table \ref{tab:r-ci-merged}).

Table \ref{tab:r-ci-merged} summarizes the 95\% confidence interval estimation results for reproduction numbers across all phases, comparing R-value estimates from ODE and Petri Net models using definition-based and Next-Generation Matrix (NGM) methods.

\begin{table}[htbp]
\setlength{\tabcolsep}{3pt}
\caption{Reproduction Number 95\% Confidence Interval Estimates Across All Phases}
\label{tab:r-ci-merged}
\centering
\small
\begin{tabularx}{\textwidth}{@{} >{\centering\arraybackslash}p{1.8cm} >{\centering\arraybackslash}p{3.5cm} >{\centering\arraybackslash}p{3.5cm} >{\centering\arraybackslash}p{3.5cm} @{} }
\toprule
\textbf{Phase} & \textbf{ODE (Definition)} & \textbf{ODE (NGM)} & \textbf{PN (NGM)}\\
\midrule
Initial & 14.67 [11.30, 20.82] & 14.67 [11.29, 20.81] & 13.90 [12.59, 15.30]\\
Phase 1 & 7.85 [6.07, 10.93] & 7.84 [6.07, 10.92] & 7.86 [7.07, 8.65]\\
Phase 2 & 7.59 [5.86, 10.45] & 7.59 [5.85, 10.45] & 7.56 [6.85, 8.35]\\
Phase 3 & 0.059 [0.046, 0.082] & 0.059 [0.046, 0.082] & 0.059 [0.053, 0.065]\\
\bottomrule
\end{tabularx}
\end{table}

\subsection{Sensitivity Analysis}

This study evaluates parameter impacts through three key metrics: sensitivity coefficients, correlation coefficients, and Sobol indices.

\begin{table}[htbp]
\setlength{\tabcolsep}{3pt}
\caption{Parameter Sensitivity Coefficient Comparison}
\label{tab:sensitivity_coefficients}
\centering
\small
\begin{tabularx}{\textwidth}{@{} >{\centering\arraybackslash}p{1.5cm} >{\centering\arraybackslash}p{2.5cm} >{\centering\arraybackslash}p{2.0cm} >{\centering\arraybackslash}p{2.0cm} >{\centering\arraybackslash}p{2.0cm} @{} }
\toprule
\textbf{Parameter} & \textbf{Parameter Description} & \textbf{ODE (Definition)} & \textbf{ODE (NGM)} & \textbf{PN (NGM)} \\
\midrule
$p$ & Symptomatic probability & 0.5543 & 0.5543 & 0.5543 \\
$\sigma$ & Latent period recovery rate & 0.0000 & 0.0001 & 0.0000 \\
$\gamma$ & Recovery rate & -1.0189 & -1.0188 & -1.0189 \\
$\epsilon$ & Relative asymptomatic infectivity & 0.1112 & 0.1112 & 0.1112 \\
$q$ & Chronicity probability & 0.0000 & 0.0000 & 0.0000 \\
$\omega$ & Chronic recovery rate & 0.0000 & 0.0000 & 0.0000 \\
$\mu$ & Mortality rate & 0.0000 & -0.0003 & 0.0000 \\
\bottomrule
\end{tabularx}
\end{table}

\begin{table}[htbp]
\setlength{\tabcolsep}{3pt}
\caption{Parameter-R Value Correlation Analysis}
\label{tab:correlation_analysis}
\centering
\small
\begin{tabularx}{\textwidth}{@{} >{\centering\arraybackslash}p{1.5cm} >{\centering\arraybackslash}p{2.5cm} >{\centering\arraybackslash}p{2.0cm} >{\centering\arraybackslash}p{2.0cm} >{\centering\arraybackslash}p{2.0cm} @{} }
\toprule
\textbf{Parameter} & \textbf{Parameter Description} & \textbf{ODE (Definition)} & \textbf{ODE (NGM)} & \textbf{PN (NGM)} \\
\midrule
$p$ & Symptomatic probability & 1.0000 & 1.0000 & 1.0000 \\
$\gamma$ & Recovery rate & -0.9142 & -0.9142 & -0.9142 \\
$\epsilon$ & Relative asymptomatic infectivity & 1.0000 & 1.0000 & 1.0000 \\
\bottomrule
\end{tabularx}
\end{table}

The recovery rate $\gamma$ emerges as the most sensitive parameter (sensitivity coefficient = -1.0189), exhibiting a strong negative correlation with $R_0$. The symptomatic probability $p$ and relative asymptomatic infectivity $\epsilon$ demonstrate moderate sensitivity with coefficients of 0.5543 and 0.1112, respectively. Other parameters exhibit negligible influence, indicating that enhancing recovery rates represents the most effective intervention measure, followed by controlling symptomatic transmission and adjusting asymptomatic infectivity.

To further quantify the global impact of parameters on $R_0$, this study also computed Sobol sensitivity indices. Table \ref{tab:sobol_sensitivity} presents the Sobol indices from multi-parameter sensitivity analysis, revealing that $\gamma$ accounts for 96.72\% of $R_0$ variance, $p$ accounts for 7.20\%, and $\epsilon$ accounts for 5.32\%. The remaining parameters contribute negligibly (<0.1\%).

\begin{table}[htbp]
\setlength{\tabcolsep}{3pt}
\caption{Sobol Global Sensitivity Index Analysis}
\label{tab:sobol_sensitivity}
\centering
\small
\begin{tabularx}{\textwidth}{@{} >{\centering\arraybackslash}p{1.5cm} >{\centering\arraybackslash}p{2.5cm} >{\centering\arraybackslash}p{2.0cm} >{\centering\arraybackslash}p{2.0cm} >{\centering\arraybackslash}p{2.0cm} @{} }
\toprule
\textbf{Parameter} & \textbf{Parameter Description} & \textbf{ODE (Definition)} & \textbf{ODE (NGM)} & \textbf{PN (NGM)} \\
\midrule
$\gamma$ & Recovery rate & 0.9672 & 0.9672 & 0.9672 \\
$p$ & Symptomatic probability & 0.0720 & 0.0719 & 0.0719 \\
$\epsilon$ & Relative asymptomatic infectivity & 0.0532 & 0.0532 & 0.0533 \\
$\mu$ & Mortality rate & 0.0007 & 0.0004 & 0.0007 \\
$\sigma$ & Latent period recovery rate & 0.0003 & 0.0004 & 0.0003 \\
$\omega$ & Chronic recovery rate & 0.0002 & 0.0003 & 0.0002 \\
$q$ & Chronicity probability & 0.0000 & 0.0000 & 0.0000 \\
\bottomrule
\end{tabularx}
\end{table}

\begin{figure}[htbp]
\centering
    \includegraphics[width=\textwidth]{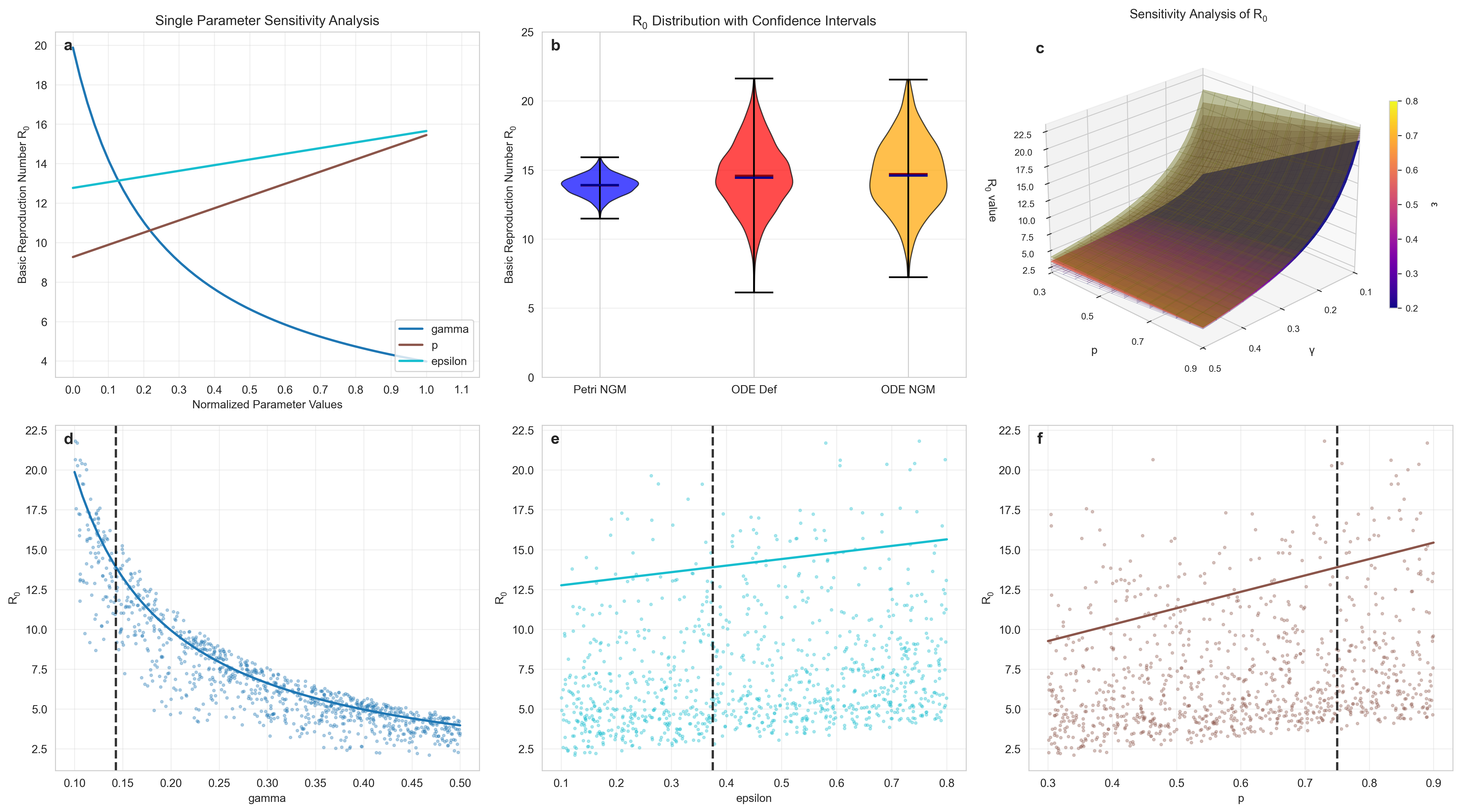}
    \caption{Visualization of parameter sensitivity analysis results. (a) Single-parameter sensitivity analysis shows the effects of recovery rate ($\gamma$), symptomatic probability ($p$), and relative asymptomatic infectivity ($\epsilon$) on the basic reproduction number ($R_0$). (b) Distribution and confidence intervals of $R_0$ values under different computational methods. (c) Three-dimensional sensitivity surface for recovery rate ($\gamma$) and symptomatic probability ($p$) versus $R_0$. (d-f) Scatter plots with trend lines for $\gamma$, $\epsilon$, and $p$, respectively.}
    \label{fig:sensitivity_analysis_plots}
\end{figure}

Figure \ref{fig:sensitivity_analysis_plots} provides intuitive validation of the quantitative results through six subplots. Subplot (a) demonstrates that the $\gamma$ curve is steepest with non-linear decline, indicating high sensitivity of $R_0$ to $\gamma$. In contrast, $p$ and $\epsilon$ display linear gentle increases with lower sensitivity, while the $p$ curve exhibits a steeper linear gradient than the $\epsilon$ curve. Subplot (b) presents violin plots showing that the three methodologies yield similar mean $R_0$ values, yet reveal distinct distribution patterns and confidence intervals. Subplot (c) displays a three-dimensional surface demonstrating maximum $R_0$ values when $\gamma$ is low and $p$ is high, with $R_0$ rapidly decreasing as $\gamma$ increases or $p$ decreases, thereby providing a multidimensional perspective on parameter interactions. Subplots (d-f) integrate single-parameter sensitivity analysis with multi-parameter Monte Carlo sampling results: curves represent systematic $R_0$ calculations at 50 parameter value points using Petri Net Next-Generation Matrix methods, while scatter points denote experimental data from 1000 multi-parameter Monte Carlo sampling runs with randomly assigned parameter values. These scatter plots confirm the quantitative analytical conclusions that $\gamma$ exhibits strong negative correlation with $R_0$, while $\epsilon$ and $p$ show moderate positive correlations with $R_0$.

In conclusion, the recovery rate $\gamma$ represents the core parameter for controlling Chikungunya transmission. Therefore, intervention strategies aimed at accelerating patient recovery will yield the most significant epidemic control effects.

%%%%%%%%%%%%%%%%%%%%%%%%%%%%%%%%%%%%%%%%%%
\section{Discussion}
\label{sec:discussion}

This study successfully constructed and validated a dual-model framework integrating Ordinary Differential Equations (ODE) and Petri Nets (PN) for rapid simulation of Chikungunya (CHIKV) transmission dynamics. Applying this framework to the 2025 Foshan Chikunganya epidemic data, we conducted phased intervention effectiveness assessment and employed two distinct methodologies (definition-based method and next-generation matrix method) to estimate basic reproduction number ($R_0$) and effective reproduction number ($R_{\mathrm{eff}}$). Results demonstrate that both models achieve comparable accuracy, with maximum R-value differences not exceeding 5.3\%. However, uncertainty analysis reveals that Petri Net models exhibit narrower 95\% confidence intervals across all phases, with initial phase CI widths of 2.71 (PN) versus 9.52 (ODE), indicating reduced parameter uncertainty in discrete event modeling.

Our analysis demonstrates that Foshan's three-phase intervention measures proved effective, particularly in the third phase where all computational methods indicated effective reproduction number $R_{\mathrm{eff}}$ had dropped below 1, successfully controlling epidemic spread. This conclusion exhibits high consistency across different models (ODE and PN) and computational methods (definition-based and NGM), enhancing result reliability. For instance, under given transmission rate $\beta(t)$, the $R$ values calculated by both methods differed minimally (0.029\%), demonstrating the internal consistency and robustness of our models.

Nevertheless, the models exhibit notable limitations. Predicted total case numbers (7,825 cases) significantly exceed observed totals (4,806 cases), consistent with systematic negative deviations observed in residual analysis (predicted values exceeding observed values), suggesting systematic overestimation during epidemic decline phases. This may stem from inadequate consideration of transmission-mitigating factors such as spontaneous population behavioral changes or data reporting delays. Despite these limitations, both models successfully captured core epidemic dynamics at macroscopic levels and quantified intervention effectiveness, providing valuable reference for public health decision-making.

In epidemiological modeling, our study contrasts and complements existing literature across several key dimensions. Regarding model frameworks, while studies such as \citep{segovia2025petri} and \citep{connolly2022from} have explored Petri Net applications in epidemiology, most research focuses primarily on theoretical framework construction or single-model validation. Our study presents the first systematic numerical comparison between ODE and PN models, validating methodological consistency through real epidemic data, which holds significant methodological innovation value.

Concerning reproduction number computational methods, our dual-method approach (definition-based method and next-generation matrix method) echoes the methodological framework of studies such as \citep{vanden2002reproduction} and \citep{guo2022computing}, yet our research further quantifies different parameter impacts on reproduction numbers through sensitivity analysis, providing more comprehensive assessment of parameter uncertainty.

Regarding public health management, sensitivity analysis results provide crucial quantitative foundations for epidemic prevention strategy formulation. Research reveals that recovery rate ($\gamma$) represents the most sensitive parameter affecting basic reproduction number $R_0$, with Sobol index reaching 0.9672, indicating this parameter's variation explains 96.72\% of total $R_0$ variation. This suggests that any measures capable of shortening disease duration and accelerating patient recovery would constitute the most effective epidemic control interventions. In contrast, symptomatic probability ($p$) and relative asymptomatic infectivity ($\epsilon$), while exhibiting positive correlations with $R_0$, demonstrate far lower sensitivity than recovery rate. This finding emphasizes the importance of optimizing clinical management and providing adequate medical resources to promote rapid patient recovery in public health practice.

Despite significant methodological and empirical analytical advances achieved in this study, several limitations require careful consideration in result interpretation and future applications. The most prominent limitation lies in the absence of detailed vector surveillance data. As a vector-borne disease, Chikungunya transmission dynamics heavily depend on key parameters including mosquito population density, infection rates, and biting behavior. Being unable to access mosquito surveillance data from Foshan's 2025 epidemic period, this study resorted to a simplified human population model, deriving asymptomatic infectious relative infectivity $\epsilon$ from empirical data. While this simplification avoided model over-parameterization issues, it may inadequately capture CHIKV transmission complexity. Additionally, although sensitivity analysis was conducted, estimates of key parameters such as relative asymptomatic infectivity ($\epsilon$) and chronicity probability ($q$) rely primarily on limited empirical data or literature values. Such uncertainty may affect model prediction reliability, particularly during policy scenario analysis.

Considering these limitations, cautious interpretation of research results is necessary. While sensitivity analysis reveals recovery rate $\gamma$ as the most sensitive parameter, a finding robust mathematically, practical application requires acknowledging that although theoretically enhancing recovery rates can significantly reduce $R_0$, in actual public health practice, recovery rates remain constrained by medical resources, treatment modalities, and disease natural course limitations, potentially limiting intervention feasibility.

In conclusion, this study's ODE-Petri Net dual-model framework not only provides reliable methodological foundations for Chikungunya epidemic modeling but also reveals key epidemic control insights through sensitivity analysis. The high consistency of model validation results ensures research conclusion credibility, while sensitivity analysis points toward optimal directions for future intervention strategies.

%% Declarations and Abbreviations
\section*{CRediT Authorship contribution statement}
Conceptualization, H.L.; methodology, H.L., J.T. and H.Y.; software, H.L.; validation, H.L., J.T. and Y.Z.; formal analysis, H.L. and H.Y.; investigation, H.L.; resources, H.L., D.C. and Y.L.; data curation, H.L.; writing---original draft preparation, J.T. and H.L.; writing---review and editing, J.T. and H.L.; visualization, H.L.; supervision, D.C., H.Y. and Z.Y.; project administration, J.T., D.C. and Y.L.; funding acquisition, D.C. and Z.Y. All authors have read and agreed to the published version of the manuscript.

\section*{Funding}
This research was funded by the Major Project of Guangzhou National Laboratory (Grant No. GZNL2024A01004) and the Science and Technology Development Fund of Macau SAR (Grant No. FDCT 0002/2024/RDP).

\section*{Ethics approval}
Not applicable.

\section*{Informed consent}
Not applicable.

\section*{Data availability}
All data and code used in this study are publicly available in the GitHub repository: \url{https://github.com/Hylouis233/ODE-Petri-Chikungunya}. Python version and dependencies are listed in the requirements file.

\section*{Acknowledgments}
We thank the Foshan Center for Disease Control and Prevention for providing valuable epidemiological data and research inspiration that enabled this study.

\section*{Declaration of competing interest}
The authors declare no conflicts of interest.

\section*{Abbreviations}
The following abbreviations are used in this manuscript:

\noindent
\begin{tabular}{@{}ll}
CHIKV & Chikungunya virus\\
ODE & Ordinary Differential Equation\\
PN & Petri Net\\
NGM & Next-Generation Matrix\\
RMSE & Root Mean Squared Error\\
MAE & Mean Absolute Error\\
MAPE & Mean Absolute Percentage Error\\
CI & Confidence Interval\\
DFE & Disease-Free Equilibrium\\
SEICR & Susceptible-Exposed-Infectious-Chronic-Recovered\\
SIR & Susceptible-Infectious-Recovered\\
MGDrivE & Mosquito Genetics Drive Evolution\\
SARS & Severe Acute Respiratory Syndrome\\
CSV & Comma-Separated Values\\
HSV & Hue-Saturation-Value\\
GPN & Generalized Petri Net\\
SAR & Special Administrative Region\\
CDC & Centers for Disease Control and Prevention\\
WHO & World Health Organization\\
AI & Artificial Intelligence
\end{tabular}

\bibliographystyle{elsarticle-harv}
\bibliography{references}
\end{document}